# How Digital Asset Treasury Companies Can Survive Bear Markets: The Case of the Strategy and Bitcoin

Hongzhe Wen, Washington University in St. Louis, hongzhewen01@gmail.com

## Abstract

Digital Asset Treasury (DAT) companies, public firms that hold large crypto reserves as a core strategy, deliver levered exposure to digital assets but face acute downside risk when equity premia over net asset value multiples (mNAV) compress in bear markets. This paper develops a survival framework that couples conservative treasury policy with an operating line that monetizes holdings independent of mark-to-market gains. Using Strategy (formerly MicroStrategy) as a case, we propose a "BTC-to-sats" payments rail that allocates a small, risk-capped liquidity sleeve of the treasury to Lightning Network channels, generating price-agnostic fee revenue (acquiring bps, routing, hedge/FX spread) while keeping settlement exposure near zero beta to BTC. We formalize a no-forced-sale condition and show how disclosed KPIs allow investors to test whether operating cash flows can bridge a 18 to 24 month bear without liquidations. The feasibility of the rail is supported by Strategy's Lightning initiative and empirical Lightning performance. Our model generalizes across DAT types and provides implementable disclosures that can sustain an mNAV premium through cycles.

# 1. Introduction

Digital Asset Treasury (DAT) companies are a recent phenomenon in corporate finance, referring to publicly traded firms that hold significant cryptocurrency reserves as a core business strategy. Pioneered by MicroStrategy (rebranded as "Strategy, Inc.") in 2020 when its CEO Michael Saylor redirected $250 million of corporate cash into Bitcoin [1], this bold approach gained rapid popularity. By 2025, over 200 public companies collectively held more than 1 million BTC (about 5% of Bitcoin's supply) on their balance sheets, with Strategy alone controlling about 640,000 BTC [2]. These DAT companies offer stock investors indirect exposure to crypto assets and were initially lauded as innovative vehicles for "alternative treasury" management and inflation hedging.

However, the crypto market's notorious volatility has put the DAT model to a harsh test. In bull markets, many DAT stocks soared far above the value of their underlying holdings as speculative investors chased upside leverage. In bear markets, though, the same leverage works in reverse: falling crypto prices can trigger severe equity drawdowns, liquidity crises, or even insolvency for imprudent firms. This paper examines how DAT companies can survive such crypto winters by adopting sustainable business models. In particular, we explore a case study of Strategy and propose how it could leverage its vast Bitcoin treasury in payment applications, effectively splitting its BTC into satoshis (small units of Bitcoin) to generate revenue. We then generalize a framework for DAT firms to endure bear markets by coupling strong treasury management with productive use cases for their digital assets.

# 2. The Rise of Digital Asset Treasury Companies

The emergence of DAT companies in the early 2020s marked a convergence of corporate finance with the crypto economy. Traditional companies like Strategy, Tesla, and Square (now Block, Inc.) began allocating portions of their treasury into Bitcoin, inspired by the idea of Bitcoin as "digital gold" or superior cash reserve. Strategy's headline-making Bitcoin buys in 2020 set the template: raise capital (via equity or debt) and turn it into a reserve of BTC, turning the company's stock into a proxy for Bitcoin itself, which provided an alternative way for investors to purchase BTC when there was no ETF in the market [2]. By 2021–2022, a wave of small-cap firms pivoted to this strategy. For example, a Japanese nail salon, a cannabis retailer, and a marketing agency all abruptly rebranded themselves as crypto holding companies, causing their stocks to skyrocket on hype alone [3]. Investors during bullish times embraced these "Bitcoin treasury" stocks as a way to gain amplified exposure to crypto within the familiar equity framework.

This craze culminated in 2025, when dozens of new DAT firms flooded the market and the sector reached "fever pitch," with some stocks spiking over 1,000% in days. Collectively, DAT companies managed over $100 billion in digital assets by mid-2025

[4], spanning Bitcoin-focused treasuries and those holding Ether, Solana, and even meme tokens. In theory, a well-run DAT could provide amplified returns in a crypto bull market by trading at an equity premium to its net asset value (NAV), essentially leveraging its holdings. Indeed, many DAT stocks initially traded at large multiples of their underlying crypto NAV (sometimes mNAV > 2 or higher) as optimism and FOMO inflated valuations. As shown in Table 1, 4 out of top 10 BTC holding companies are traded at mNAV > 2 or higher. This premium, alongside access to capital markets, enabled some firms to continuously issue new shares and buy even more crypto, forming a feedback loop during euphoric conditions.

| Ticker | Company | BTC held | Mkt Cap | mNAV | BTC/Share |
|---|---|---|---|---|---|
| MSTR | Strategy | 640,808 | $77,395M | 1.095 | 0.00223146 |
| CEP | XXI | 43,514 | $4,895M | 1.020 | 0.00016278 |
| MTPLF | Metaplanet | 30,823 | $3,663M | 1.077 | 0.00002701 |
| CEPO | Bitcoin Standard Treasury Company | 30,021 | $215M | 0.065 | 0.00146444 |
| BLSH | Bullish | 24,300 | $5,725M | 2.136 | 0.00021464 |
| DJT | Trump Media & Technology Group | 15,000 | $3,686M | 2.227 | 0.00006237 |
| CLSK | CleanSpark | 13,011 | $5,003M | 3.486 | 0.00004629 |
| TSLA | Tesla | 11,509 | $1,472,611M | 1159.877 | 0.00000357 |
| GDC | GD Culture Group | 7,500 | $74M | 0.090 | 0.00044655 |
| CANG | Cango | 6,394 | $691M | 0.979 | 0.00003712 |

Table 1: Top 10 publicly traded BTC holding companies (cryptocurrency related companies excluded) [5].

Yet, as history shows, such hype-driven premiums are hard to sustain, even for Strategy. The DAT model's growth fueled itself on investor enthusiasm, but it also imported all the risks of crypto volatility onto corporate balance sheets. By late 2025, sentiment had shifted: analysts began warning that many DAT companies were overleveraged imitators of Strategy's bold bet [6]. Once the bear hits, most of the tokens are facing substantial dips in price, which might lead to serious consequences. The crucial question is: Were these companies structurally prepared to handle a market downturn, or had they simply ridden a one-way wave of speculative fervor?

## 3. Challenges and Risks in the DAT Model

The core risk for a digital asset treasury company is that its stock price becomes tightly coupled to the volatile market value of its crypto holdings. Essentially, a DAT stock is a high-beta bet on the underlying asset (Bitcoin or others). When crypto prices soar, the stock can overshoot to the upside; but when crypto enters a bear market, the firm's market capitalization can plunge even more dramatically. This was evidenced in 2022–2023's "crypto winter," when Bitcoin fell ~75% from its peak: Strategy's own stock (MSTR) cratered by a similar magnitude as investors questioned its leverage and durability. Moreover, ever since Strategy started to purchase BTC, its stock price shows a correlation of 94.321% with the price of BTC, which indicates an extremely high correlation between the two.

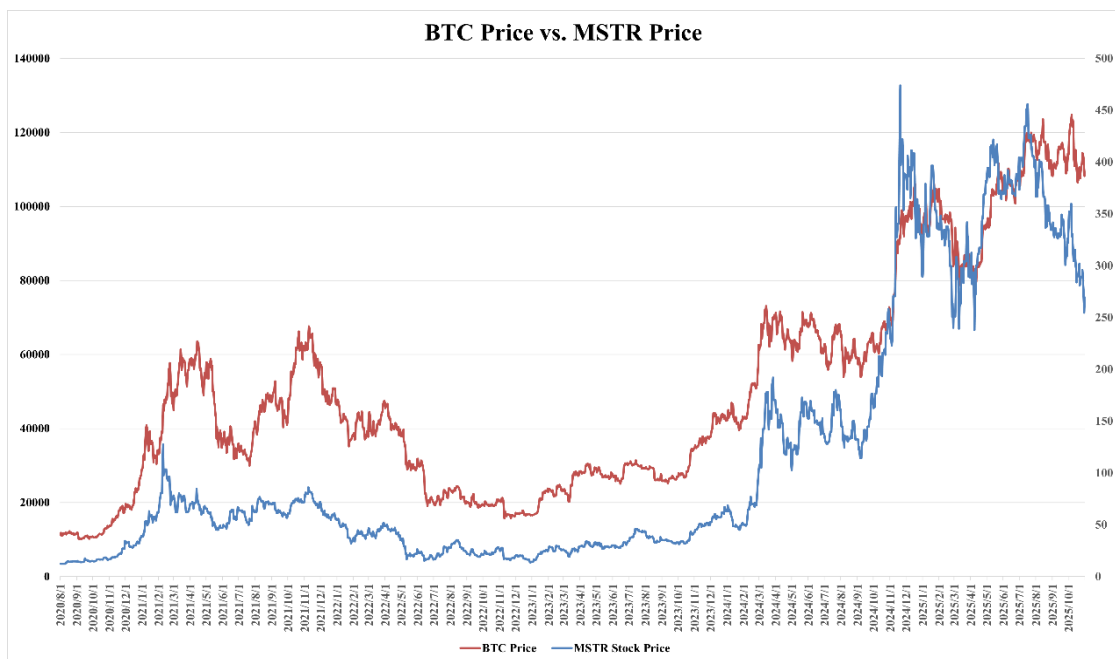

Figure 1: BTC price vs MSTR stock price (data from Yahoo Finance and CoinGecko).

Many smaller imitators fared worse. By Q3 2025, at least five DAT companies' stocks (e.g. MetaPlanet, Sharplink Gaming, Ton Strategy, ETHZilla, etc.) traded below the value of the Bitcoin on their balance sheets, meaning the market implied their operating business was worthless or even negative. This inversion, a market cap under NAV, is an existential threat to the DAT model. As the Financial Times observed, these firms "need to trade above their underlying crypto assets. Otherwise, they would not be able to follow Saylor's strategy and keep purchasing crypto [7]." In response, several had to borrow money (up to $250M in some cases) just to buy back their own shares in hopes of boosting the price. Such defensive buybacks are a telling sign of distress, described as potentially the "death rattle" of the DAT craze.

Moreover, many DAT companies undertook their crypto accumulation with significant leverage or dilution. Some issued convertible bonds or took loans to buy cryptocurrencies; others continuously sold new equity (including private investments in public equity (PIPE) deals) to raise funds. These tactics work in a rising market but can backfire in a decline. New equity issuance at depressed prices alienates shareholders (due to dilution) and often leads to further price drops. Debt, meanwhile, introduces fixed obligations that can become hard to meet if crypto prices fall (since the firm has little other income).

An example was Metaplanet, a Japanese Bitcoin-holding firm: its market NAV multiple (mNAV) fell below 1.0 during a market crash (indicating its stock < NAV), prompting the CEO to consider buybacks and even issuing preferred shares to restructure capital [8]. The underlying peril is that a heavily indebted or dilutive DAT in a bear market might be forced to liquidate holdings at the worst time, further depressing crypto prices in a vicious cycle.

Another challenge is asset diversification and liquidity. Some DATs deviated from the Bitcoin-focused playbook, betting on altcoins or more illiquid tokens to entice investors. These multi-asset treasuries generally carried higher risk: smaller cap tokens can evaporate in value or become illiquid in a downturn. Indeed, 2025 saw cases of DAT firms buying Dogecoin (CleanCore Solutions), BNB, even novelty tokens, only to see their stocks plunge when those bets soured.

It is notable that Bitcoin is uniquely suited as a corporate treasury asset due to its large scale, liquidity, and emerging institutional infrastructure (ETFs, custodians, derivatives). Companies focusing on obscure crypto assets may face a "liquidity vacuum" in bear markets, unable to sell or hedge positions without crashing the price. Many such firms were effectively taking on the high risk of altcoins under the guise of a treasury strategy.

In summary, the DAT model's risks include: extreme price-volatility impact on equity, dependence on continued investor confidence (premium to NAV) to raise capital, potential for leverage-induced failures, and asset-specific liquidity problems. Many DATs lack meaningful cash flows and rely solely on the market value or yield of their crypto holdings, making them structurally fragile when the tide turns. Notably, in September 2025, Semler Scientific, a struggling Bitcoin holder, agreed to be acquired by an asset manager, effectively rescuing it with a 210% premium buyout [9]. Consolidation like this suggests that weaker treasury companies either find a savior or face collapse. Clearly, not every company that rode the Bitcoin boom will survive the bust.

## 4. Key Strategies for Survival in Bear Markets

In essence, strong governance, prudent financial management, and real profitability are the key of survival. Below, we outline key strategies a DAT firm should employ:

## 4.1 Conservative Treasury Management and Liquidity Reserves

A resilient DAT anticipates volatility and plans for downturns. This means holding enough cash or stable assets to cover operating expenses and debt service for multiple years, so the company is not forced into fire-sales of crypto at cycle bottoms. It also means avoiding short-term debt or margin loans that could trigger margin calls. Strategy, for instance, largely used long-dated convertible bonds for its Bitcoin buys and maintained substantial Bitcoin collateral; this allowed it to avoid forced liquidation even when BTC dropped over 50% in 2022. In contrast, firms that took on short-term, high-interest loans against their crypto (or, worse, used leverage on exchanges) found themselves in precarious positions.

## 4.2 Discipline in Capital Structure

When a DAT's stock tumbles, panicked moves like selling equity at rock-bottom prices or issuing dilutive PIPE deals can destroy shareholder trust. Investors can accept volatility in Bitcoin's price; what they reject is management diluting their stake or scrambling for emergency funds in a downturn. To maintain confidence, companies should raise capital proactively during good times (when valuations are high) and then pause during bad times.

In practice, this might mean cutting or freezing further crypto purchases in a bear market, rather than issuing cheap shares to keep buying. It might also involve creative financing that aligns with long-term investors (e.g. issuing preferred stock or bonds when conditions are favorable, instead of desperation moves later). The overarching principle is capital discipline, which do not overextend when the market is frothy, and do not overreact when the market is fearful.

## 4.3 Real Business Operations

The most important survival factor is having a source of income independent of crypto price appreciation. A sustainable DAT is one that is a normal profitable business first, and a crypto holder second. Companies that issued stock solely to buy crypto (with no profitable core business) face the valid question of why investors should not just hold the crypto directly. By contrast, a company that actually earns revenue (e.g. from product sales or services) can continually buy and hold Bitcoin out of its profits, which is far more sustainable. Bitcoin in that scenario is a reserve asset or long-term investment, not the sole source of "growth."

Strategy's own CFO has repeatedly stated that despite Bitcoin volatility, the company's software business covers its interest payments and operating costs, providing a baseline stability [10]. A profitable underlying business allows a DAT to survive the crypto winter without selling off assets. It also justifies a continuing premium to NAV, since the company offers value beyond just holding coins.

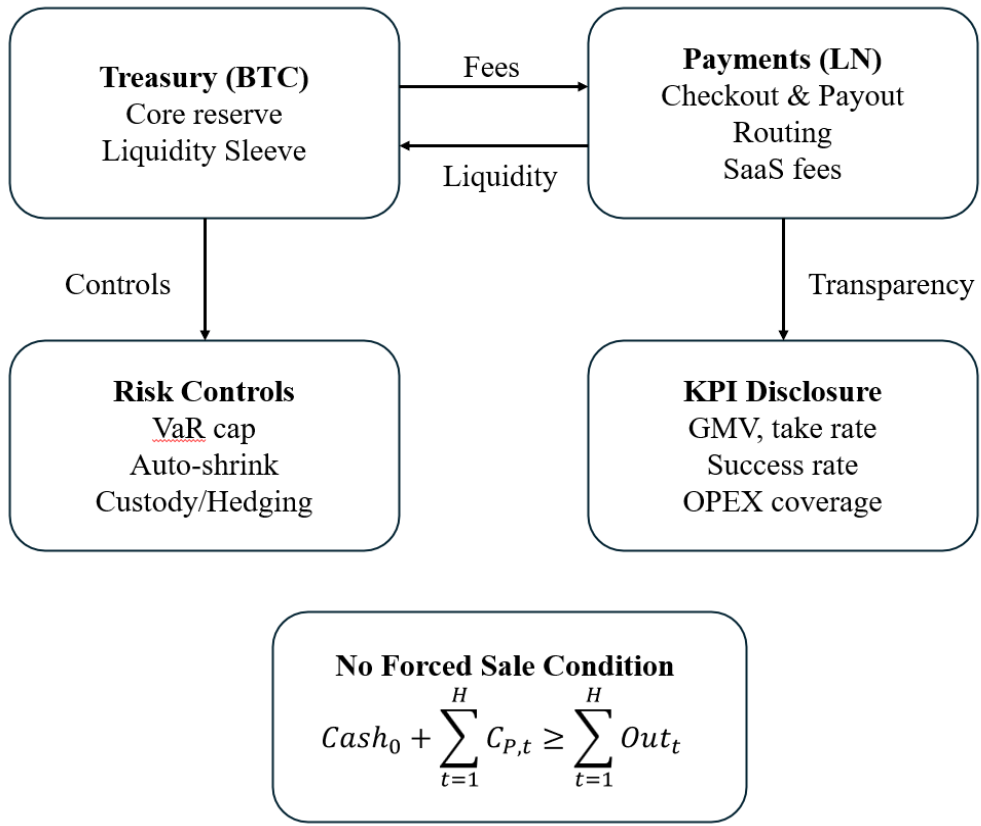

Figure 3: Dual-Engine DAT Model: Treasury (T) + Payments (P).

## 4.4 Governance and Strategic Clarity

Strong governance entails transparent, well-communicated strategy and alignment with shareholders' interests. Strategy's governance could be seen in Saylor's personal conviction (he dramatically increased his own stake, aligning with shareholders) and in splitting roles so that a dedicated CEO could run the software business while Saylor focused on Bitcoin strategy. Companies must articulate to investors why their holding of crypto adds value and how they will handle downturns. For example, some firms have outlined policies like: no crypto sales unless required, or a willingness to pause acquisitions if premiums vanish.

Good governance also means strong security and compliance practices (to avoid devastating hacks or regulatory sanctions). Cases of poorly secured treasuries or dubious related-party dealings would quickly unravel investor confidence. In essence, DAT executives must manage their crypto reserves with the same prudence and transparency expected of any corporate treasury.

## 4.5 Asset Focus and Risk Management

Another strategic choice is which crypto assets to hold and how to hedge or utilize them. Bitcoin's depth and increasing institutional adoption make it behave more like a macro asset (with established derivatives for hedging), whereas niche tokens can suffer catastrophic declines or vanish. Thus, a survival-oriented DAT will likely concentrate on Bitcoin (and possibly secondarily on top assets like Ether with caution). If a company does diversify, it should be extremely judicious.

For instance, engaging in staking for yield on some ETH holdings might be reasonable (earning protocol rewards to supplement income), but speculative forays into illiquid DeFi coins or using DeFi leverage can be an action that collapses in a market disruption.

The better approach is risk-managed use of assets: hedging a portion of holdings with options or futures during clear downtrends, or allocating a minority of reserves to stablecoins or cash equivalents to buffer against volatility. While many Bitcoin treasuries (Strategy included) chose not to hedge (preferring full upside exposure), from a survival standpoint having at least some downside insurance or diversification in stable assets can be wise, especially if the firm has debt to service. The key is that any such hedging or yield strategy must be transparent and limited to avoid the fate of over-engineered financial bets. Simple balance sheet with mostly BTC and some cash, and possibly modest hedges, is easier for investors to understand and trust than one filled with obscure tokens and leverage.

Implementing these strategies can allow a DAT company to endure a prolonged bear market. In practical terms, a well-run DAT should be able to say: "Even if our crypto assets drop 50-80% in value and capital markets shut to us for 2 years, we can still pay all our bills, avoid forced asset sales, and continue building our business."

## 5. Case Study: Strategy's Bitcoin-to-Satoshis Payment Business Model

Strategy provides a valuable case study for how a DAT company can evolve its strategy to be more bear-resistant. As the first and largest Bitcoin treasury company, Strategy has navigated both steep bull runs and brutal bears, learning important lessons along the way. By 2025, the company held approximately 628,000 BTC (over 3% of total supply) with an aggregate cost basis around $46 billion. This massive position gives Strategy huge upside in a crypto rally. Indeed, when Bitcoin's price doubled in early 2025, Strategy reported quarterly net income of $10 billion largely from unrealized gains. But it also exposes the firm to downside: a 50% drop in BTC would theoretically wipe out tens of billions in asset value.

Strategy survived the 2022 bear market in part due to long-term debt structure and Saylor's unwavering "HODL" philosophy (the company steadfastly refused to sell off

Bitcoin, aside from a one-time minor sale for tax advantages). Yet, even Strategy faced market skepticism. In August 2025, its stock briefly lost the premium over its BTC holdings, trading near 1:1 with NAV, undermining its ability to raise fresh equity. To sustain its strategy in the long run, Strategy recognized it needed to do more than just buy and hold Bitcoin; it needed to leverage those Bitcoin in business operations to generate cash flow and justify its valuation.

Michael Saylor's answer was to actively use Bitcoin. Starting in 2022, Strategy launched an initiative to build products on the Bitcoin Lightning Network, a layer-2 protocol that enables fast, low-cost Bitcoin transactions (measured in satoshis, the smallest BTC units) [11].

In practical terms, Strategy's vision was to deploy some of its Bitcoin into Lightning Network channels and develop enterprise software that allows millions of people to transact in satoshis [12]. For example, one idea Saylor floated was an online business giving promotional rewards to new customers who sign up for a service or fill out a survey using satoshis. Another idea was to have paywalls or premium content access secured by Lightning payments, where users pay a few sats to read an article or access an API. These use-cases would showcase Bitcoin's capability as a payment medium and create new revenue streams for Strategy (either by charging clients for the software or by taking a small fee per transaction).

By 2023, Strategy had begun rolling out a concrete product: the Lightning Rewards platform. This is a software-as-a-service offering that lets companies reward their customers or employees with Bitcoin -payments (satoshis) via the Lightning Network [13]. According to Strategy's website, the Lightning Rewards platform supports virtually unlimited scalability and integrates with popular enterprise systems like Salesforce, soft 365, Zoom, and more.

Companies can customize reward rules and the platform handles the instant distribution of those satoshi rewards globally, at negligible cost. Crucially, it provides an administrative portal with analytics to track engagement and measure ROI of these Bitcoin incentives. In effect, Strategy has taken its treasury asset (BTC) and built a payments and loyalty solution around it, one that did not exist in traditional finance.

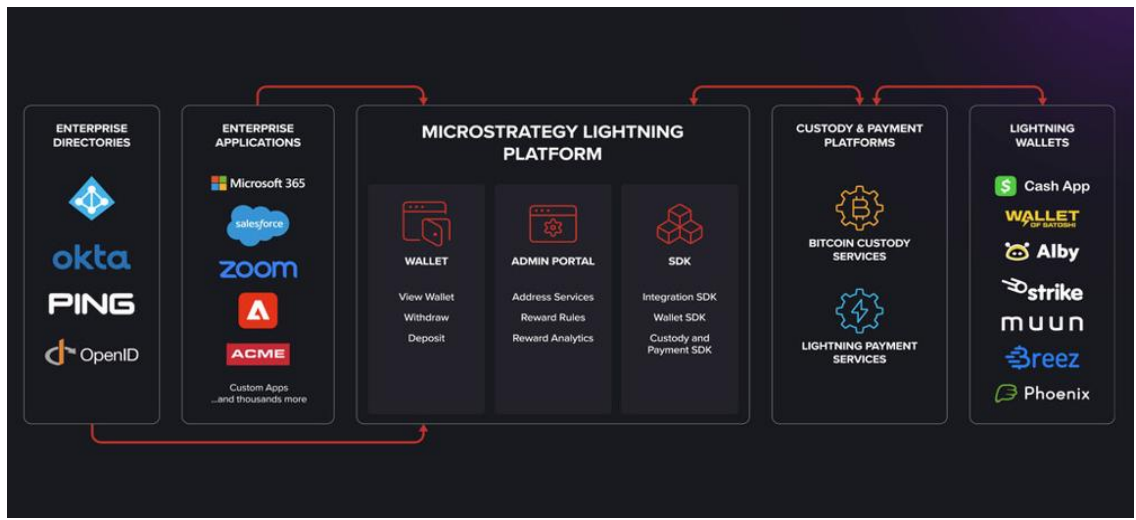

Figure 4: Architecture of Strategy's Lightning Network platform (Lightning Rewards). The platform integrates with enterprise directories (Okta, etc.) and applications (Salesforce, Zoom, etc.), providing a Bitcoin Lightning wallet and admin portal. It connects to Lightning payment services and popular Lightning wallets (Cash App, Strike, Phoenix, etc.), enabling companies to reward or transact with users in Bitcoin (satoshis) instantly and globally [66].

As shown in Figure 4, the Strategy Lightning Platform is designed to slot into a corporate IT environment seamlessly. Employees or customers receive a Lightning wallet (web and mobile) through which they can earn, view, and withdraw their satoshi rewards. Additionally, Strategy enabled Lightning addresses that look like email addresses (e.g. saylor@strategy.com) to simplify receiving payments. This email integration, done in collaboration with a Lightning service called Bottlepay, was a novel demonstration: anyone could send sats to Saylor's company email, showing how user-friendly Bitcoin payments could become. For 1 satoshi (worth only fractions of a cent), the Lightning Network can process up to a million transactions per second, highlighting the scalability and negligible cost of this layer-2 solution. By embracing these technologies, Strategy aims to monetize its Bitcoin holdings beyond passive appreciation, essentially turning its corporate treasury into a productive asset.

The Lightning Rewards platform offers Strategy multiple potential revenue streams:

First, it makes the company's analytics software offering more compelling to enterprises looking for modern incentive programs; this could drive software licensing or subscription fees. Second, Strategy can charge usage fees or earn a spread on Bitcoin it channels through the Lightning Network for clients. Third, by pioneering business use-cases for Bitcoin, Strategy strengthens the Bitcoin ecosystem, which indirectly supports the value of its own holdings, a virtuous cycle if successful. Saylor indicated that Strategy's goal is to help any enterprise "spin up Lightning wallets for 100,000 employees or 10 million customers overnight." [14] The company has dedicated R&D teams on Lightning enterprise applications, including wallet, server, and authentication systems.

Importantly, Strategy's core business before Bitcoin was enterprise analytics software, and that remains an ongoing (if modest) source of revenue. In Q2 2025, its software business pulled in about $115 million (with growing cloud subscription revenues) [15].

While small relative to its Bitcoin-related gains, this recurring revenue base provides stability. The Lightning Network initiative smartly bridges the two sides of Strategy's business: it uses the company's Bitcoin expertise to create a new software product, thus marrying the treasury strategy with operational income. In essence, Strategy is morphing from simply a Bitcoin holding company into a Bitcoin-powered software company.

This evolution could prove critical in a bear market. If Bitcoin's price slumps for an extended period, Strategy can lean on its Lightning services to generate cash (or at least greatly enhance its value proposition to customers), reducing the need to liquidate BTC. It also differentiates Strategy from a hypothetical Bitcoin ETF, where an ETF can hold BTC but cannot develop business solutions or earn revenues from using BTC. By developing a real payment business, Strategy provides a rationale for investors to hold MSTR stock beyond just "Bitcoin tracking."

However, there are challenges and risks: Lightning Network itself, while growing, is still a nascent technology with evolving liquidity and some UX hurdles. Strategy is essentially betting that Lightning will achieve wide adoption, which is not guaranteed. Additionally, competitors (including startups focused solely on Lightning services) could vie for the same enterprise clients.

## 6. From Sats Rewards to a Sats Payments Rail

Strategy's Lightning Rewards demonstrates that satoshi-denominated payouts are technically and operationally feasible at enterprise scale. The natural next step is to extend this capability from incentives to a general payments rail that merchants, platforms, and payroll programs can use daily. The business logic is straightforward: repurpose a small, bounded sleeve of the firm's Bitcoin treasury to provide Lightning liquidity (opening and maintaining channels), then monetize payment acquiring fees and modest routing fees while keeping settlement flows economically hedged.

In practice, the sats rail offers three features:

1. Merchant Checkout: Customers pay in sats; merchants receive either BTC or fiat (T+0/T+1). Price is locked at quote and hedged in the background.

2. Global Payouts/Payroll: Mass payouts to creators or contractors in sats with optional local fiat off-ramps.

3. Embedded Wallets & Sats-Back: White-label Lightning wallets for apps, plus sats-back at checkout (funded by the merchant using part of the fee savings).

For the usage of treasury, Strategy allocates a liquidity sleeve (for example, 2–5% of BTC holdings) to open channels and maintain high success rates for payments. This sleeve is managed with strict risk limits: channels shrink automatically in stress, and routing/float exposure is capped in both size and time (seconds to minutes). The core treasury remains long-term and untouched.

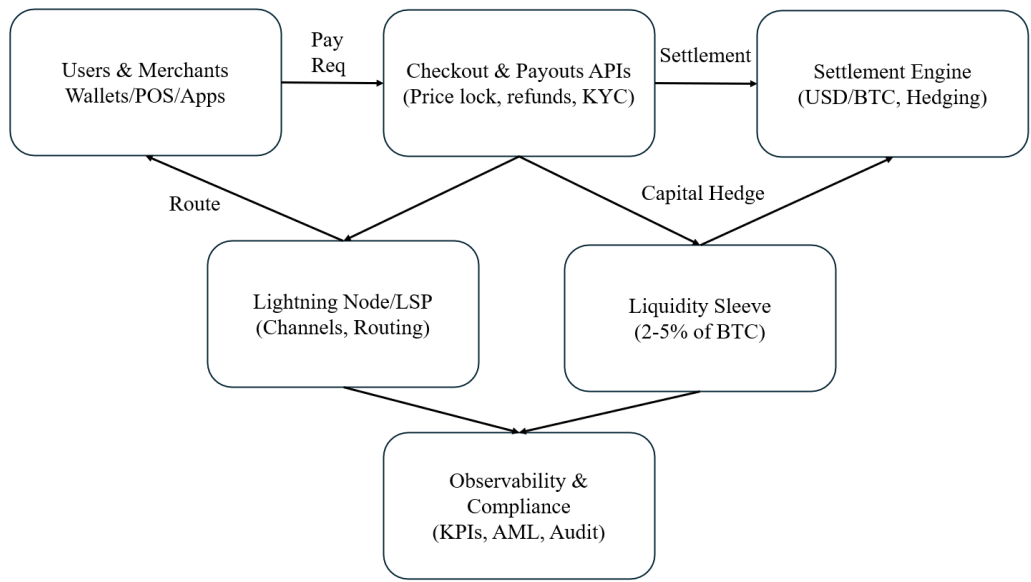

Figure 4: Satoshi Payments Rail Architecture (Conceptual)

The revenue stack is intentionally plain:

- Acquiring take rate: ~10–50 bps on processed volume (GMV), tiered for enterprise.

- Hedge/FX spread (optional): a few bps when settling to fiat.

- Routing fees: tiny per-payment, but meaningful at scale.
  Because Lightning removes card interchange and many chargeback costs, merchants can often fund sats-back rewards and still lower total acceptance cost versus cards (typically 200–300 bps).

The rail runs under standard money-movement controls (KYC/AML, sanctions screening, consumer disclosures) with fair-value accounting for held BTC and back-to-back hedge accounting on fiat settlements. The point is not regulatory arbitrage, but cost, speed, and programmability.

At the same time, Strategy should disclose a small set of operating KPIs so investors can separate operating progress from BTC price moves: GMV, take rate, payment success rate, routing revenue per 100k transactions (tx), rebalancing cost (bps), merchant churn, and, critically, the share of operation expenses (OPEX) covered by non-mark-to-market fee revenue.

The rail's role in a downturn is to help the company avoid becoming a forced seller of BTC. Let monthly operating outflows (OPEX + interest + maintenance capex) be $Out_t$, and non-mark-to-market payments cash inflow be $C_{P,t}$. With an initial cash reserve $Cash_0$ and horizon H months, the no-forced-sale condition is:

$$Cash_0 + \sum_{t=1}^{H} C_{P,t} \geq \sum_{t=1}^{H} Out_t$$

Operationally, Strategy targets H = 18 – 24 months, hedges settlement inventory to keep payment P&L near zero beta to BTC, and sets a value-at-risk cap on the liquidity sleeve (for example, ≤ 20% of cash reserves). If this inequality holds under modeled bear paths (e.g., −70% BTC), Strategy can carry full core BTC exposure through the cycle, preserving upside convexity into the next expansion.

Moving from rewards to a payments rail converts a portion of the treasury's latent optionality into recurring, price-agnostic fee revenue. That revenue, verified by transparent KPIs, supports a persistent premium over NAV in public markets and provides a concrete, cash-flow-based buffer in bear markets without liquidating the core BTC position.

## 7. Toward a Sustainable DAT Business Model

Drawing from the Strategy case and broader industry observations, we can outline a generalized business model for DAT companies aimed at long-term sustainability (especially through bear markets). The crux of the model is to balance two roles:

1. Asset Manager: stewarding a portfolio of digital assets (e.g. Bitcoin) with prudent financial practices.

2. Service Provider: leveraging those assets to generate revenue or business value.

Below are the key components of this model:

### 7.1 Prudent Treasury Management

A sustainable DAT treats its crypto treasury like a strategic reserve, not a speculative trading account. This means maintaining a conservative capital structure (low debt, especially low short-term debt), and keeping sufficient non-crypto liquidity. For example, a company might decide to always keep one to two years of operating expenses in cash or stablecoins, separate from its crypto holdings. This acts as a buffer during crypto downturns. Stress testing should guide what percentage drawdown the firm can withstand.

If not, the company should adjust its capital structure by raising equity during bear to fortify cash reserves, or by limiting how much of its balance sheet is in crypto. In practice, it may also involve basic hedging: a DAT might purchase put options or insurance that pay out if crypto prices crash below a certain level, ensuring funds for debt or critical expenses, which avoids the company to become a forced seller of the treasury asset at the worst moment.

## 7.2 Core Business Operations and Profitability

The DAT model works best as an extension of an existing profitable enterprise, rather than a standalone bet. Thus, companies should focus on excelling in their primary business domain (be it software, fintech, mining, payments, etc.) and use crypto as an enhancement. For instance, a Bitcoin mining company naturally holds Bitcoin, but its survival still depends on efficient mining operations and cost management. Similarly, a tech company that holds crypto should continue to grow its software or services revenue, using crypto to complement rather than replace its income.

The "Bitcoin as a treasury asset" narrative is more compelling when it is a portion of a successful business's balance sheet (much like Apple holding cash or gold) than when it is the entire story. This implies DAT firms might need to invest in developing real products or services if they currently have none. For example, if a company holds a lot of Ether, perhaps it can run validator nodes and offer staking-as-a-service to other investors, earning fees. If it holds Bitcoin, it could offer Bitcoin-backed loans or integrate Lightning payments (similar to Strategy's approach). These ancillary services turn the treasury from a passive asset into a part of the revenue engine.

## 7.3 Supporting the Crypto Ecosystem (Ecosystem Revenue)

DAT business model could include plans to engage more deeply with the ecosystems of their chosen assets. A Bitcoin-focused company might invest in Bitcoin infrastructure, develop tools, or provide liquidity that strengthens Bitcoin's network (like payments, security, or institutional access), thereby finding new revenue sources. We see this with Strategy's Lightning platform and also with companies like Block (Square) which integrated Bitcoin buying and Lightning transfers into Cash App, generating transaction fees. An Ethereum-focused DAT might similarly participate in DeFi infrastructure or enterprise blockchain solutions (provided it manages the risks).

The idea is that by adding value to the network you are invested in, you not only potentially earn income, but also building the adoption and utility of that network, which could positively impact the asset's long-term value. In effect, the company becomes a stakeholder that is paid for helping the ecosystem grow. This strategy can justify the company's existence beyond just holding coins. They might become known for expertise or services in that crypto domain (e.g. "Company X is not just a large Ether holder, but also runs major staking pools and provides smart contract auditing services"). During bear markets, other participants might still pay for those services, keeping the company afloat.

## 7.4 Flexible Asset Utilization (Yield and Lending)

Another component is finding relatively low-risk ways to put treasury assets to work to earn yield. Simply holding an asset like Bitcoin yields nothing. But a DAT could allocate a portion of holdings to conservative yield-generating activities. For example, some might lend out a fraction of their Bitcoin to reputable institutional borrowers or on regulated platforms to earn interest (with proper collateral and risk management). Others might use Bitcoin as collateral to borrow fiat at low interest, providing operating capital without selling the Bitcoin, using crypto to fund expansion and then repaying when conditions improve.

Staking is applicable to proof-of-stake assets: a company holding large amounts of a PoS coin (like SOL or ETH) can stake it to earn protocol rewards (often 5-10% annually). The caution here is that chasing high yields can lead to excessive risk due to the asset that is used for leverage-on-leverage yield farming can turn into a nothing in a crisis. So the model would only include moderate, well-understood yield tactics: e.g., stake vanilla ETH, but do not leverage it; lend some Bitcoin, but only over-collateralized to blue-chip borrowers. The earned yield (even if just 2-5% annually) provides some cash inflow during bears, helping cover salaries or overhead without selling assets. Over time, compounded yield can also increase the treasury size in a self-sustaining way.

## 7.5 Preparedness for Regulatory and Accounting Changes

Lastly, a viable model must anticipate the evolving regulatory landscape. As crypto becomes more mainstream, regulators may impose new rules on corporate holdings, disclosures, or even capital requirements for crypto exposure. Companies should get ahead by implementing accounting and audit practices for digital assets.

DATs should also obtain any necessary licenses (for example, if they start offering payment services or custody, they might need money transmitter licenses or chartered subsidiaries). Being proactive here avoids forced changes later that could disrupt the business. It also confers legitimacy that might make institutional investors more comfortable owning the stock. In short: treat the crypto treasury with the seriousness of a financial asset – secure it (multi-signature, custody solutions), insure it if possible, audit it, and comply with any emerging best practices or regulations. Companies that do so will stand out as stable and trustworthy in an often Wild West sector.

By combining the above elements, a DAT company transforms from a pure asset play into a hybrid entity: part investment fund, part operating business. The theoretical outcome is a company whose valuation is not solely a direct multiple of its crypto holdings, but also factors in cash flows and growth prospects from business lines built on those holdings. In a bear market scenario, such a company would have multiple levers to pull: it can tighten its belt on expenses, lean on its non-crypto revenue, earn a bit of yield from its crypto, maybe even gain market share as weaker competitors fold.

It would not need to liquidate its Bitcoin at the bottom, meaning when the cycle turns bullish again, it still holds its full stash to ride the upside.

It is worth noting that not every DAT from the 2021–2025 wave will be able to implement all these changes in time. We will likely see a natural selection process. Those survivors will effectively be the ones that followed the principles discussed. In many ways, this maturation mirrors the early dot-com era: after the bubble, only companies with sound business models (and sometimes new models learned through adversity) survived to become long-term winners.

## 8. Conclusion

Digital Asset Treasury companies represent an innovative convergence of corporate strategy and decentralized finance. The dramatic rise and fall of many DAT stocks between 2020 and 2025 underscore both the potential and peril of this model. On one hand, a company like Strategy showed that accumulating a large Bitcoin treasury could transform its fortunes, aligning it with one of the best-performing assets of the decade. On the other hand, the subsequent volatility revealed that simply holding crypto is not a panacea for corporate growth, which companies must earn their premium by adding value beyond just HODLing. The experiences of this period have yielded a clear lesson: to survive the bear markets, DAT firms must evolve from passive asset holders into active, value-generating enterprises.

Strategy's journey exemplifies this evolution. By doubling down on Bitcoin but also finding ways to deploy it (via Lightning payments and rewards), it set a blueprint for how to make a crypto treasury strategy resilient. It demonstrated that Bitcoin can be both a store of value on the balance sheet and the basis of new business ventures that produce income. This dual approach, reserve asset plus revenue asset, may become the gold standard for DAT companies going forward.

Looking ahead, the concept of digital asset treasuries is likely to remain, but it will be a more nuanced and mature incarnation. We may see fewer companies attempting this model, but those that do will be larger, more professionally managed, and integrated into the crypto economy in substantive ways. Bitcoin's growing acceptance (with ETFs, major custodians, etc.) means holding BTC as a reserve asset could become almost as normal as holding foreign currency or gold. When that happens, the novelty premium will fade and only real operational excellence will set companies apart.

In conclusion, the era of reckless crypto treasury plays is winding down, and a new era is beginning, one where digital asset treasuries are managed with the same rigor as any corporate fund, and where these assets are actively employed to advance the company's mission. A bear market, rather than a mortal threat, can then be seen as a proving ground that tempers and strengthens these firms. By adhering to sound financial principles and innovating at the intersection of blockchain technology and business, DAT companies can indeed survive the winter and emerge ready to blossom in the next spring of the crypto markets.